%% file: main_2025_08_26.tex
\documentclass[letterpaper,conference]{ieeeconf}
\IEEEoverridecommandlockouts                              
\usepackage[utf8]{inputenc}
\usepackage{textcomp}

\usepackage{graphicx}
\usepackage{amsmath}
\usepackage[version=4]{mhchem}
\usepackage{siunitx}
\usepackage{graphicx}
\usepackage{longtable,tabularx}
\usepackage[style=ieee,doi=false]{biblatex}
\usepackage{tikz}
\usepackage{pgfplots}
\pgfplotsset{compat=1.18}

\usepackage[table]{xcolor}
\usepackage{array}
\usepackage{comment}
\usepackage{xcolor} 
\newcolumntype{Y}{>{\centering\arraybackslash}X}

\usetikzlibrary{positioning, arrows.meta}
\input{PackagesCommands}

\usepackage[colorinlistoftodos,textwidth=1 in,textsize=footnotesize]{todonotes}

\DeclareMathOperator\atanh{atanh}

\title{Backstepping Control of a Bicopter with Unknown Mass and Inertia}

\title{Globally Asymptotically Stable Adaptive Control of a \\ 
Bicopter with Unknown Mass and Inertia}

\title{Estimation inversion-free Adaptive Control of a \\ 
Bicopter with Unknown Mass and Inertia}

\title{Singularity-free Backstepping-based Adaptive Control of a \\ 
Bicopter with Unknown Mass and Inertia}
\title{Singularity-free Adaptive Control of a \\ 
Bicopter with Unknown Mass and Inertia}

\title{Stability Preserving and Constraint Enforcing Control of a Bicopter}
\title{Stability Preserving Safe Control of a Bicopter}

\author{
Jhon Manuel Portella Delgado
and
Ankit Goel
\thanks{Jhon Manuel Portella Delgado and Ankit Goel are with the Department of Mechanical Engineering, University of Maryland, Baltimore County, 1000 Hilltop Circle, Baltimore, MD 21250. {\tt\small jportella, ankgoel@umbc.edu}}%
}

\begin{document}

\maketitle




\begin{abstract}
This paper presents a control law for stabilization and trajectory tracking of a multicopter subject to safety constraints. 
The proposed approach guarantees forward invariance of a prescribed safety set while ensuring smooth tracking performance.
Unlike conventional control barrier function methods, the constrained control problem is transformed into an unconstrained one using state-dependent mappings together with carefully constructed Lyapunov functions. 
This approach enables explicit synthesis of the control law, instead of requiring a solution of constrained optimization at each step.
The transformation also enables the controller to enforce safety without sacrificing stability or performance.
Simulation results for a polytopic reference trajectory confined within a designated safe region demonstrate the effectiveness of the proposed method.

%
%
%
%
%
\end{abstract}
\textit{\bf keywords:} state constraints, stability, multicopter.

\section{Introduction}


    




Multicopters are increasingly employed in precision agriculture \cite{mukherjee2019}, environmental monitoring \cite{lucieer2014,klemas2015}, construction \cite{li2019}, and cargo transport \cite{villa2020}.
Despite their widespread adoption, reliable and safe control remains challenging due to nonlinear, time-varying dynamics and uncertain operating conditions.
Safety-critical requirements in control system design are typically formulated as state constraints, which ensure that key states remain within designated safe sets \cite{ames2019cbf}. 
However, the nonlinear and underactuated nature of multicopter dynamics makes simultaneous enforcement of safety constraints and closed-loop stability an open challenge.
Research on constrained control of multicopters has explored model predictive control (MPC) \cite{bemporad2009hierarchical,nguyen2021model}, barrier Lyapunov functions (BLFs) \cite{li2021finite}, and control barrier functions (CBFs) in cascaded loops \cite{khan2020barrier}.
MPC enforces constraints over a prediction horizon, but it is challenging to apply to multicopters due to nonlinear, underactuated dynamics.
Linearization to obtain quadratic programs introduces approximation errors \cite{wang2020}, while solving the full nonlinear problem is computationally demanding and often infeasible on embedded hardware \cite{SCHLAGENHAUF20207026}, limiting MPC in fast maneuvers.
BLFs use Lyapunov functions that grow unbounded near constraint boundaries, but singular terms require delicate gain tuning, and constraints can only be imposed on transformed error states rather than directly on position and velocity \cite{Sadeghzadeh2023}.
CBFs have gained traction for ensuring forward invariance of safe sets, but they are not stand-alone controllers—requiring integration with nominal controllers and optimization to enforce Nagumo’s condition \cite{nagumo1942lage}.
Stability is not generally guaranteed \cite{Kim2023}, and constructing higher-order CBFs to enforce position and velocity constraints simultaneously remains challenging.


This paper develops a controller for a multicopter that enforces both position and velocity constraints while guaranteeing closed-loop stability.
To keep the mathematical construction simple and emphasize the control synthesis procedure, we focus on a bicopter system, which is a multicopter constrained to a vertical plane and thus is modeled by a 6th-order nonlinear system instead of a 12th-order nonlinear system. 
Despite the lower dimension of the state space, the 6th-order bicopter retains the complexities of the nonlinear and underactuated dynamics of an unconstrained multicopter.
The adaptive control of a bicopter system was considered in our prior work reported in \cite{Portella2024fixed,portella2024adaptive}. 


The proposed control synthesis approach guarantees stability through Lyapunov theory and circumvents the singularities associated with barrier Lyapunov functions by applying a state transformation that converts the constrained problem into an unconstrained one.
In this formulation, the bounded safe set is first mapped to an unbounded domain.
For clarity of exposition, symmetric bounds are assumed; however, with a suitable choice of transformation, the method can be easily extended to asymmetric bounds.
Within the backstepping framework, a controller is then designed in the transformed state space.
A composite Lyapunov candidate—constructed from a logarithmic term and a hyperbolic cosine term—is used.
This function exhibits quadratic behavior near the origin and linear growth away from it.
Consequently, its slope approaches a constant at large values, a key property that enables the enforcement of velocity constraints.
The main contributions of this work include the introduction of a state transformation that converts the constrained control problem into an unconstrained one, the novel use of a log-cosh function in the construction of the Lyapunov function, and the development of an intermediate control law that guarantees the semi-definiteness of the Lyapunov function’s time derivative.





The paper is organized as follows. 
Section \ref{sec:prob_formulation} reviews the bicopter dynamics.
Section \ref{sec:MCBC} presents the stability-preserving safe control synthesis and its stability analysis. 
Section \ref{sec:simulations} presents the results of numerical simulations to validate the proposed controller.
Finally, the paper concludes with a discussion of results and future research directions in section \ref{sec:conclusions}.

\section{Bicopter Dynamics}
\label{sec:prob_formulation}

\begin{figure}[h]
    \centering
    \resizebox{0.75\columnwidth}{!}{
    
\tikzset{every picture/.style={line width=0.75pt}} 

\begin{tikzpicture}[x=0.75pt,y=0.75pt,yscale=-0.7,xscale=0.7]

\draw   (382.73,198.06) .. controls (384.44,196.15) and (387.38,195.98) .. (389.3,197.69) .. controls (391.22,199.41) and (391.38,202.35) .. (389.67,204.26) .. controls (387.96,206.18) and (385.02,206.35) .. (383.1,204.63) .. controls (381.18,202.92) and (381.02,199.98) .. (382.73,198.06) -- cycle ;
\draw    (385.37,137.84) -- (385.88,195.97) ;
\draw [shift={(385.35,135.84)}, rotate = 89.5] [fill={rgb, 255:red, 0; green, 0; blue, 0 }  ][line width=0.08]  [draw opacity=0] (12,-3) -- (0,0) -- (12,3) -- cycle    ;
\draw    (449.36,199.25) -- (391.23,200.02) ;
\draw [shift={(451.36,199.23)}, rotate = 179.24] [fill={rgb, 255:red, 0; green, 0; blue, 0 }  ][line width=0.08]  [draw opacity=0] (12,-3) -- (0,0) -- (12,3) -- cycle    ;
\draw  [fill={rgb, 255:red, 0; green, 0; blue, 0 }  ,fill opacity=1 ] (384.93,201.16) .. controls (384.93,200.46) and (385.5,199.9) .. (386.2,199.9) .. controls (386.9,199.9) and (387.47,200.46) .. (387.47,201.16) .. controls (387.47,201.86) and (386.9,202.43) .. (386.2,202.43) .. controls (385.5,202.43) and (384.93,201.86) .. (384.93,201.16) -- cycle ;
\draw    (322.35,83.75) -- (220.35,195.75) ;
\draw [shift={(271.35,139.75)}, rotate = 132.32] [color={rgb, 255:red, 0; green, 0; blue, 0 }  ][fill={rgb, 255:red, 0; green, 0; blue, 0 }  ][line width=0.75]      (0, 0) circle [x radius= 1.34, y radius= 1.34]   ;
\draw    (200.75,177.75) -- (220.35,195.75) ;
\draw   (183.73,197.05) .. controls (181.5,194.8) and (183.84,188.88) .. (188.95,183.81) .. controls (194.05,178.75) and (200,176.46) .. (202.22,178.71) .. controls (204.45,180.95) and (202.11,186.88) .. (197.01,191.94) .. controls (191.9,197.01) and (185.95,199.29) .. (183.73,197.05) -- cycle ;
\draw   (202.22,178.71) .. controls (200,176.46) and (202.33,170.54) .. (207.44,165.47) .. controls (212.55,160.41) and (218.49,158.12) .. (220.72,160.37) .. controls (222.94,162.61) and (220.61,168.54) .. (215.5,173.6) .. controls (210.39,178.66) and (204.45,180.95) .. (202.22,178.71) -- cycle ;
\draw    (302.75,65.75) -- (322.35,83.75) ;
\draw   (285.73,85.05) .. controls (283.5,82.8) and (285.84,76.88) .. (290.95,71.81) .. controls (296.05,66.75) and (302,64.46) .. (304.22,66.71) .. controls (306.45,68.95) and (304.11,74.88) .. (299.01,79.94) .. controls (293.9,85.01) and (287.95,87.29) .. (285.73,85.05) -- cycle ;
\draw   (304.22,66.71) .. controls (302,64.46) and (304.33,58.54) .. (309.44,53.47) .. controls (314.55,48.41) and (320.49,46.12) .. (322.72,48.37) .. controls (324.94,50.61) and (322.61,56.54) .. (317.5,61.6) .. controls (312.39,66.66) and (306.45,68.95) .. (304.22,66.71) -- cycle ;
\draw  [dash pattern={on 4.5pt off 4.5pt}]  (331.49,138.96) -- (271.35,139.75) ;
\draw  [fill={rgb, 255:red, 0; green, 0; blue, 0 }  ,fill opacity=1 ] (326.07,201.55) .. controls (326.07,200.98) and (326.53,200.51) .. (327.1,200.51) .. controls (327.68,200.51) and (328.14,200.98) .. (328.14,201.55) .. controls (328.14,202.12) and (327.68,202.59) .. (327.1,202.59) .. controls (326.53,202.59) and (326.07,202.12) .. (326.07,201.55) -- cycle ;
\draw [line width=1.5]    (410.42,70.59) -- (410.12,109.92) ;
\draw [shift={(410.09,113.92)}, rotate = 270.44] [fill={rgb, 255:red, 0; green, 0; blue, 0 }  ][line width=0.08]  [draw opacity=0] (8.75,-4.2) -- (0,0) -- (8.75,4.2) -- (5.81,0) -- cycle    ;
\draw [line width=1.5]    (196.22,172.04) -- (172.07,149.28) ;
\draw [shift={(169.16,146.54)}, rotate = 43.29] [fill={rgb, 255:red, 0; green, 0; blue, 0 }  ][line width=0.08]  [draw opacity=0] (8.75,-4.2) -- (0,0) -- (8.75,4.2) -- (5.81,0) -- cycle    ;
\draw [line width=1.5]    (298.42,60.42) -- (274.27,37.66) ;
\draw [shift={(271.35,34.92)}, rotate = 43.29] [fill={rgb, 255:red, 0; green, 0; blue, 0 }  ][line width=0.08]  [draw opacity=0] (8.75,-4.2) -- (0,0) -- (8.75,4.2) -- (5.81,0) -- cycle    ;
\draw   (192.75,92.55) .. controls (192.54,89.99) and (194.46,87.75) .. (197.02,87.55) .. controls (199.58,87.35) and (201.82,89.26) .. (202.02,91.83) .. controls (202.22,94.39) and (200.31,96.63) .. (197.75,96.83) .. controls (195.18,97.03) and (192.95,95.11) .. (192.75,92.55) -- cycle ;
\draw    (151.07,49) -- (193.41,88.83) ;
\draw [shift={(149.61,47.63)}, rotate = 43.25] [fill={rgb, 255:red, 0; green, 0; blue, 0 }  ][line width=0.08]  [draw opacity=0] (12,-3) -- (0,0) -- (12,3) -- cycle    ;
\draw    (239.68,45.24) -- (200.04,87.77) ;
\draw [shift={(241.05,43.78)}, rotate = 132.99] [fill={rgb, 255:red, 0; green, 0; blue, 0 }  ][line width=0.08]  [draw opacity=0] (12,-3) -- (0,0) -- (12,3) -- cycle    ;
\draw  [fill={rgb, 255:red, 0; green, 0; blue, 0 }  ,fill opacity=1 ] (196.51,93.1) .. controls (196,92.62) and (195.99,91.82) .. (196.47,91.31) .. controls (196.95,90.81) and (197.75,90.79) .. (198.26,91.27) .. controls (198.77,91.76) and (198.78,92.56) .. (198.3,93.07) .. controls (197.82,93.57) and (197.01,93.59) .. (196.51,93.1) -- cycle ;
\draw  [draw opacity=0] (295.51,113.84) .. controls (295.56,113.84) and (295.62,113.84) .. (295.67,113.84) .. controls (301.93,113.99) and (306.85,120.97) .. (306.66,129.45) .. controls (306.57,133.33) and (305.43,136.85) .. (303.61,139.52) -- (295.32,129.19) -- cycle ; \draw    (295.67,113.84) .. controls (301.93,113.99) and (306.85,120.97) .. (306.66,129.45) .. controls (306.59,132.56) and (305.85,135.43) .. (304.62,137.82) ; \draw [shift={(303.61,139.52)}, rotate = 291.91] [fill={rgb, 255:red, 0; green, 0; blue, 0 }  ][line width=0.08]  [draw opacity=0] (7.2,-1.8) -- (0,0) -- (7.2,1.8) -- cycle    ; \draw [shift={(295.51,113.84)}, rotate = 24.68] [fill={rgb, 255:red, 0; green, 0; blue, 0 }  ][line width=0.08]  [draw opacity=0] (7.2,-1.8) -- (0,0) -- (7.2,1.8) -- cycle    ;

\draw (413.67,79.33) node [anchor=north west][inner sep=0.75pt]   [align=left] {{\small $\vect g$}};
\draw (453,190) node [anchor=north west][inner sep=0.75pt]   [align=left] {{\small $\hat \imath _{\rm{A}}$}};
\draw (375,120) node [anchor=north west][inner sep=0.75pt]   [align=left] {{\small $\hat \jmath _{\rm{A}}$}};
\draw (234,31) node [anchor=north west][inner sep=0.75pt]  [rotate=-313.75] [align=left] {{\small $\hat \imath _{\rm B}$}};
\draw (129.9,40) node [anchor=north west][inner sep=0.75pt]  [rotate=-313.75] [align=left] {{\small $\hat \jmath _{\rm B}$}};
\draw (287,20.33) node [anchor=north west][inner sep=0.75pt]   [align=left] {{\small ${\vect f}_2$}};
\draw (184,135.33) node [anchor=north west][inner sep=0.75pt]   [align=left] {{\small ${\vect f}_1$}};
\draw (260.67,125) node [anchor=north west][inner sep=0.75pt]   [align=left] {{\small c}};
\draw (327.33,200) node [anchor=north west][inner sep=0.75pt]   [align=left] {{\small \textit{w}}};
\draw (308.33,115) node [anchor=north west][inner sep=0.75pt]   [align=left] {{\small $\theta$}};
\end{tikzpicture}
}
\caption{Bicopter configuration considered in this paper. 
The bicopter is constrained to the {$\hat  \imath _\rmA- \hat \jmath _\rmA$} plane and rotates about the {$\hat k_\rmA$} (out of the page) axis of the inertial frame {$\rm F_A.$}
$\rm F_B$ is fixed to the bicopter such that the forces $\vect f_1$ and $\vect f_2$ are along $\hat \jmath_\rmB.$
Note that $\vect r_{\rmc/w} = r_1 \hat \imath_\rmA + r_2 \hat \jmath_\rmA.$}
    \label{fig:Bicopter}
\end{figure}
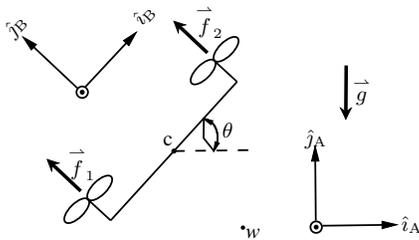

We consider the bicopter system shown in Figure \ref{fig:Bicopter} and described in detail in Section II of \cite{portella2024adaptive}.
%
The equations of motion of a bicopter are
\begin{align}
    m \ddot r_1 = -F \sin(\theta), \
    m \ddot r_2 = F \cos(\theta) - m g, \
    J \ddot \theta = M,
    \label{eq:eom_theta_2}
\end{align}
where 
$m$ and $J$ are the mass and the moment of inertia of the bicopter, respectively,  
$r_1$ and $r_2$ are the horizontal and vertical positions of the center of mass,
$\theta$ is the roll angle, 
and 
$F\isdef f_1+f_2$ and $M\isdef(f_2-f_1)\ell$ are the total force and the total moment applied to the bicopter, respectively, where $\ell$ is the length of the bicopter arm. 

As shown in our previous work in \cite{portella2024adaptive}, the backstepping framework can not be used with the bicopter system due to the non-invertibility of the input map.
In a bicopter system, the input map is a tall matrix, and thus, the non-invertibility is structural rather than due to rank deficiency at a few points in the state space. 
As shown in \cite{portella2024adaptive}, a dynamic extension of the equations of motion overcomes the structural non-invertibility of the input map. 
The derivation of the equations of motion and their dynamic extension is described in more detail in \cite{portella2024adaptive}.
%
The dynamically extended equations of motion are
\begin{align}
    \dot x_1 &= x_2,
    \label{eq:x1dot}
    \\
    \dot x_2 &= f_2 + g_2(x_3) ,
    \label{eq:x2dot}
    \\
    \dot x_3 &= x_4,
    \label{eq:x3dot}
    \\
    \dot x_4
        &=
            g_4
            u \label{eq:x4dot},
\end{align}
where
\begin{align}
    x_1 \isdef \matl
                 r_1
                 \\
                 r_2
               \matr, 
    x_2 \isdef \matl
                 \dot r_1
                 \\
                 \dot r_2
               \matr,
    x_3 \isdef \matl
                 \theta
                 \\
                 F
               \matr, 
    x_4 \isdef \matl
                 \dot \theta
                 \\
                 \dot F
               \matr,
\end{align}
\begin{align}
    u
        \isdef 
            \matl 
                \ddot F \\
                M
            \matr,
    f_2
        \isdef 
            \matl
                0\\
                -g
            \matr,
\end{align}

and the functions 
\begin{align}
    g_2(x_3) 
        &\isdef
            m\inv
            \matl
                -\rmS_{x_{31}}\\
                \rmC_{x_{31}} 
            \matr 
            x_{32}, \quad
    g_4
        \isdef 
            \matl
                0 & J^{-1}\\
                1 & 0
            \matr.
    \label{eq:bicopter_def_matrices}
\end{align}

Note that the system \eqref{eq:x1dot}-\eqref{eq:x4dot} is in pure feedback form since $x_3$ appears non-affinely in \eqref{eq:x2dot}. 
Finally, since $g_4$ is a constant nonsingular matrix, the backstepping framework can be used with the extended equations. 

The objective is to stabilize the bicopter system \eqref{eq:x1dot}-\eqref{eq:x4dot} and follow the commanded trajectory, while ensuring that
$x_1 | < \overline{x}_{1}$ and
$x_2 | < \overline{x}_{2}$ for all time,
where $\overline{x}_{1}, \overline{x}_{2} \in \BBR^{2}_+$ are the user-defined position and velocity bounds.

In set notation, the objective can formulated as the design of a control law that ensures the command is followed and the position $x_1$ and velocity $x_2$ of the bicopter satisfy $(x_1,x_2) \in \SSS,$ where the safety set $\SSS$ is 
\begin{align}
    \SSS
        &=
            \{
                (p,v) \colon 
                p,v\in \BBR^2, 
                |p_1| < \overline{x}_{11}, 
                \nn \\ &
                |p_2| < \overline{x}_{12}, 
                |v_1| < \overline{x}_{21}, 
                |v_2| < \overline{x}_{22}
            \}.
\end{align}


\section{Stability Preserving Safe Control}
\label{sec:MCBC}
In this section, we design a controller using the backstepping framework to stabilize the bicopter system subject to both position and velocity constraints within a prescribed safe set $\SSS$.
To streamline the controller construction, we assume that the bicopter’s objective is to track a constant waypoint.
Nonetheless, the proposed procedure can be readily extended to cases where the desired state trajectory is a sufficiently smooth function of time, as discussed in \cite{johnson2024continuous}, by incorporating the corresponding derivatives into the control law.

\begin{figure}[!ht]
    \centering
    \includegraphics[width=1\columnwidth]{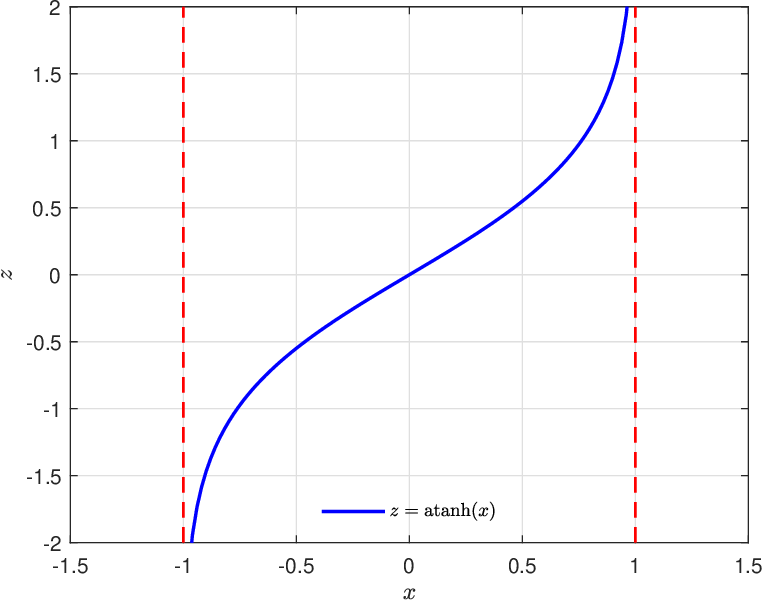}
    \caption{
    Nonlinear map $\atanh{x}$ used to map the constraint set to an unconstrained set. 
    }
    \label{fig:x_to_z_map}
\end{figure}

The control problem with position and velocity constraints is first reformulated into a problem without constraints through a state transformation, ensuring that safety limits are automatically satisfied.
In particular, the inverse hyperbolic tangent function, shown in Figure \ref{fig:x_to_z_map}, is used to transform the position and velocity states.
As shown in Figure \ref{fig:x_to_z_map}, while $x \in (-1,+1),$ the corresponding mapped value satisfies $z \in \BBR.$
Within this transformed space, a backstepping-based procedure is employed to construct the control law incrementally, using intermediate variables and Lyapunov candidate functions to guarantee stability at each stage. 
A logarithmic-hyperbolic function is introduced in the Lyapunov design to achieve both quadratic behavior near the equilibrium and linear growth away from it, enabling effective enforcement of velocity constraints. 
The resulting control law explicitly ensures that the system remains within the prescribed safe set while asymptotically converging to the desired waypoint. 
Rigorous stability analysis, based on the Barbashin–Krasovskii–LaSalle invariance principle, confirms that the proposed controller achieves both forward invariance of the safety set and asymptotic stability of the closed-loop dynamics.

\subsection{State Transformation}

Define, for $i=1,2,$
\begin{align}
    \chi_{1i} 
        &\isdef
            x_{1i}/\overline{x}_{1i}, 
    \quad 
    \chi_{2i} 
        \isdef
            x_{2i}/\overline{x}_{2i},
\end{align}
Note that the constraints are satisifed if, for $i=1,2,$ $\chi_{1i}, \chi_{2i} \in [-1,1].$
Next, define 
\begin{align}
    z_1 
        &\isdef
            D(\overline{x}_1) \atanh (\chi_{1}), 
    \label{eq:z1}
    \\
    z_2
        &\isdef
            D(\overline{x}_2) \atanh (\chi_{2}), 
    \label{eq:z2}
    \\
    z_3 
        &\isdef 
            x_3, 
    \label{eq:z3}
    \\
    z_4
        &\isdef 
            x_4,
    \label{eq:z4}
\end{align}
where, for $q \in \BBR^n,$
\begin{align}
    D(q) &\isdef {\rm diag}(q_1, \ldots , q_n) \in \BBR^{n \times n}.
\end{align}
Note that the graph of $\atanh(.)$ map is shown in Figure \ref{fig:x_to_z_map}.

Next, to simplify the expressions in the control synthesis, we define, for $i=1,2,$
\begin{align}
    \zeta_{1i} 
        &\isdef
            z_{1i}/\overline{x}_{1i}, 
    \quad 
    \zeta_{2i} 
        \isdef
            z_{2i}/\overline{x}_{2i}.
    \label{eq:zeta_def}
\end{align}
It thus follows from \eqref{eq:z1}-\eqref{eq:z4} that
\begin{align}
    x_1
        &=
            D(\overline{x}_1) \tanh(\zeta_1)
            ,
    \label{eq:x1_transformed}
    \\
    x_2
        &=
           D(\overline{x}_2) \tanh(\zeta_2)
    \label{eq:x2_transformed}
    \\
    x_3 &= z_3
    \label{eq:x3_transformed}
    \\
    x_4 &= z_4.
    \label{eq:x4_transformed}
\end{align}
Note that the constraints are satisfied for all values of $z_1, z_2 \in \BBR^2.$

The dynamics in terms of the transformed state $z$ is
\begin{align}
    \dot z_1
        &=
            \SF_1(z_1, z_2) 
        =
            D({\rm Ch}_{\zeta_1})^2 
            D(\overline{x}_2)
            \tanh (\zeta_2)
    \label{eq:z1_dot}
    \\
    \dot{z}_2
        &=
            D({\rm Ch}_{\zeta_2})^2
            \left(
                f_2
                +
                g_2(z_3) 
            \right)
    \label{eq:z2_dot}
    \\
    \dot{z}_3
        &=
            z_4
    \label{eq:z3_dot}
    \\
    \dot{z}_4
        &=
            g_4u,
    \label{eq:z4_dot}
\end{align}
where, for $q \in \mathbb{R}^n,$
\begin{align}
    {\rm Ch_q}
        \isdef
            \cosh
                \left(
                    \matl
                        q_1
                        &
                        q_2
                        &
                        \cdots
                        &
                        q_n
                    \matr^{\rmT}
                \right)
                \in \mathbb{R}^n
\end{align}

\subsection{Controller Synthesis}
Let 
$
    x_{\rmd1}
        =
        \matl 
            r_{\rmd1} & r_{\rmd2} 
        \matr^{\rmT} 
        \in \mathbb{R}^2
$ 
be the desired value of $x_1$,  
$\chi_{\rmd1}$ be the corresponding desired value of $\chi_1, $
and 
$
z_{\rmd1} 
    \isdef 
        D(\overline{x}_1)
        \atanh(\chi_{\rmd1})
        \in \BBR^2
$ 
be the corresponding desired value of $z_1.$
Define the tracking error $e_1 \isdef z_1 - z_{\rmd1}.$

\subsubsection{$e_1$ Stabilization}

Consider the function
\begin{align}
    V_1
        &\isdef
            \frac{1}{2}
            e_1^{\rmT}
            e_1.
    \label{eq:V_1}
\end{align}
Differentiating \eqref{eq:V_1} 
and using \eqref{eq:z1_dot} yields
\begin{align}
    \dot{V}_1
        =
            e_1^\rmT\dot{z}_1
                =
                    e_1^\rmT
                    \SF_1(z_1,z_2)
\end{align}
Note that if
\begin{align}
    \SF_2(z_1,z_2)
        &= 
            -k_1 e_1,
\end{align}
where $k_1>0,$ then $\dot V_1 < 0.$ 
However,
$
   \SF_1(z_1,z_2)
$ 
cannot be chosen arbitrarily.
Instead, we continue the design process as shown below to formulate a control law that ensures the desired $\SF_1(z_2,z_2)$ response. 

Define
\begin{align}
    e_2
        &\isdef
            \SF_1(z_1,z_2)
            +
            k_1
            e_1,
    \label{eq:e2_def}
\end{align}
where 
$k_1>0.$ Then, differentiating \eqref{eq:V_1} yields
\begin{align}
    \dot{V}_1
        &=
            - k_1e_1^{\rmT}e_1
            + e_2^{\rmT}e_1.
    \label{eq:V1_dot_intermidiate}
\end{align}

\subsubsection{$e_2$ Stabilization}
Consider the function
\begin{align}
    V_2
        &\isdef
            V_1
            +
            \log
            \left(
                \cosh
                    \left(
                        \zeta_{21}
                    \right)
            \right)
            +
            \log
            \left(
                \cosh
                    \left(
                        \zeta_{22}
                    \right)
            \right).
    \label{eq:V2}
\end{align}
Differentiating \eqref{eq:V2} and using \eqref{eq:z2_dot}, \eqref{eq:e2_def}, and \eqref{eq:V1_dot_intermidiate}  yields
\begin{align}
    \dot{V}_2
        &=
            - k_1e_1^{\rmT}e_1
            + e_2^{\rmT}e_1
            +
            \tanh(\zeta_2)^\rmT 
            D_\rmI(\overline{x}_2) 
            \dot{z}_2
        \nn 
        \\
        &=
            - k_1e_1^{\rmT}e_1
            + e_2^{\rmT}e_1
            +
            \left(
                e_2
                -
                k_1e_1
            \right)^{\rmT}
            Q
            \left(
                f_2
                +
                g_2(z_3) 
            \right),
            \label{eq:M2_def}
\end{align}
where
\begin{align}
    Q
        &\isdef
            D({\rm Ch}_{\zeta_2})^2
            D_\rmI({\rm Ch}_{\zeta_1})^2
            D_\rmI(\overline{x}_2)^2
            \in \BBR^{2 \times 2},
    \label{eq:M_def}
\end{align}
and
where, for $q \in \BBR^n$ such that, for $i=1,\ldots, n,$ $q_i \neq 0,$
\begin{align}
    D_\rmI(q) &\isdef {\rm diag}(q_1\inv, \ldots , q_n\inv) \in \BBR^{n \times n}.
\end{align}
Note that $Q$ is a product of three diagonal matrices with strictly positive elements, and is therefore positive definite.

Next, note that if
\begin{align}
    f_2 + g_2(z_3) = - (e_2 - k_1 e_1),
\end{align}
then $\dot V_2 < 0.$
However, $g_2(z_3)$ can not be chosen arbitrarily.
Instead, we continue the design process as shown below to formulate a control law that ensures the desired $g_2(z_3)$ response. 

Define
\begin{align}
    e_3
        &\isdef
            Q
            \left(
                f_2
                +
                g_2(z_3)
            \right)
            +
            k_2e_2,
    \label{eq:e3_def}
\end{align}
where $k_2 > 0.$
Differentiating \eqref{eq:V2} and letting $k_2 = k_1\inv $ yields
\begin{align}
    \dot{V}_2
        &=
            - (\sqrt{k_1}e_1 - \sqrt{k_2}e_2)^{\rmT}
              (\sqrt{k_1}e_1 - \sqrt{k_2}e_2)
    \nn \\
    & \quad 
            + e_3^{\rmT}
              \left(
                e_2 - k_1e_1
              \right).
    \label{eq:V2_dot_intermediate}
\end{align}
\begin{remark}
The choice $k_2 = k_1\inv$ allows the cancellation of several terms and the construction of the positive semidefinite first term in $\dot V_2$ in \eqref{eq:V2_dot_intermediate}.     
\end{remark}


%
\subsubsection{$e_3$ Stabilization}
Consider the function
\begin{align}
    V_3
        &\isdef
            V_2
            +
            \dfrac{1}{2}
            e_3^{\rmT}
            e_3.
    \label{eq:V3}
\end{align}
Differentiating \eqref{eq:V3} and using \eqref{eq:V2_dot_intermediate} yields
\begin{align}
    &\dot{V}_3
        =
            \dot V_2
            +
            e_3^\rmT \dot e_3
        \nn \\
        &=
            - (\sqrt{k_1}e_1 - \sqrt{k_2}e_2)^{\rmT}
              (\sqrt{k_1}e_1 - \sqrt{k_2}e_2)
            \nn \\ &
            +
            e_3^\rmT 
            \Big(
                 e_2 - k_1e_1
                +
                \dot Q(f_2+g_2(z_3)) 
                +
                Q N z_4 
                +
                k_2
                \dot{e}_2
            \Big),
\end{align}
where
\begin{align}
    N(z_3) \isdef \partial_{z_3} g_2(z_3) \in \BBR^{2\times 2}.
    \label{eq:N_def}
\end{align}

Note that if 
\begin{align}
     &e_2 - k_1e_1
                +
                \dot Q(f_2+g_2(z_3)) 
                \nn \\ &\quad 
                +
                Q N z_4 
                +
                k_2
                \dot{e}_2
        =
        -k_3 e_3,
\end{align}
then
\begin{align}
    \dot{V}_3
        &=
            - (\sqrt{k_1}e_1 - \sqrt{k_2}e_2)^{\rmT}
              (\sqrt{k_1}e_1 - \sqrt{k_2}e_2)
            \nn \\ &\quad 
            -k_3 e_3^\rmT e_3.
\end{align}
However, $z_4$ can not be chosen arbitrarily.
Instead, we continue the design process as shown below to formulate a control law that ensures the desired $z_4$ response. 

Define
\begin{align}
    e_4 
        &\isdef 
             e_2 - k_1e_1
                +
                \dot Q(f_2+g_2(z_3)) 
                \nn \\ &\quad 
                +
                Q N z_4 
                +
                k_2
                \dot{e}_2
                +
                k_3 e_3,
    \label{eq:e4_def}
\end{align}
where $k_3>0.$
Then, differentiating \eqref{eq:V3} yields
\begin{align}
\dot{V}_3
     &=
            - (\sqrt{k_1}e_1 - \sqrt{k_2}e_2)^{\rmT}
              (\sqrt{k_1}e_1 - \sqrt{k_2}e_2)
    \nn \\
    & \quad 
            - k_3e_3^{\rmT}e_3 + e_4^{\rmT}e_3.
    \label{eq:V3_dot_intermediate}
\end{align}

\subsubsection{$e_4$ Stabilization}
Consider the function
\begin{align}
    V
        &\isdef
            V_3
            +
            \dfrac{1}{2}e_4^{\rmT}e_4.
    \label{eq:V4}
\end{align}
Differentiating \eqref{eq:V4}  yields
\begin{align}
    \dot{V}
        &=
            \dot{V}_3
            +
            e_4^{\rmT}\dot{e}_4
    \nn \\
        &=
           - (\sqrt{k_1}e_1 - \sqrt{k_2}e_2)^{\rmT}
              (\sqrt{k_1}e_1 - \sqrt{k_2}e_2)
    \nn\\
    \quad &
           - k_3e_3^{\rmT}e_3
           + e_4^{\rmT}
           \left(
                \Phi
                +
                \Psi
                u
           \right),
    \label{eq:V_dot}
\end{align}
where
\begin{align}
    \Phi
        &\isdef
            e_3
            +
    \dot{e}_2
                -
                k_1
                \dot{z}_1
                +
                \ddot{Q}
                \left(
                    f_2
                    +
                    g_2(z)_3
                \right)
            + k_2
              \ddot{e}_2
    \nn \\
    \quad &
            + k_3
                \dot{e}_3
            +
            2\,
            \dot{Q}
            Nz_4
            +
            \dot N
            z_4,
    \label{eq:Phi_def}
    \\
    \Psi
        &\isdef
            Q
            N
            g_4.
            \label{eq:beta_def}
\end{align}

Finally, consider the control law
\begin{align}
    u
        &=
            -
            \Psi^{-1}
            \left(
                \Phi
                +
                k_4 e_4
            \right),
    \label{eq:control_law}
\end{align}
where
$k_4 > 0.$
Substituting the controller \eqref{eq:control_law} in \eqref{eq:V_dot} yields
\begin{align}
    \dot{V}
        &=
           - (\sqrt{k_1}e_1 - \sqrt{k_2}e_2)^{\rmT}
              (\sqrt{k_1}e_1 - \sqrt{k_2}e_2)
    \nn\\
    \quad &
           - k_3e_3^{\rmT}e_3
           - k_4e_4^{\rmT}e_4.
    \label{eq:V_dot_final}
\end{align}


\begin{remark}
    Note that the control law in \eqref{eq:control_law} requires $\Psi$ given by \eqref{eq:beta_def} to be nonsingular. 
Since $Q$ given by \eqref{eq:M_def} and $g_4$ given by \eqref{eq:bicopter_def_matrices} 
are positive definite, $\Psi$ is nonsingular if and only if $N(z_3)$ is nonsingular. 
As shown in \eqref{eq:partial_g2_z3}, $N(z_3)$ is nonsingular if and only if 
$z_{32} = x_{32} = F \neq 0.$    
\end{remark}

\begin{remark}
    The control law \eqref{eq:control_law} requires $\Psi$ and $\Phi,$ which require $\dot e_2, \dot z_1, \dot Q,\ddot Q,$ and $\dot N.$
    These derivatives are not obtained through differentiation, but are explicitly computed as functions of the measured state, as shown in Appendix \ref{ap:math_expressions}.
    
\end{remark}

\begin{remark}

Although the control \eqref{eq:control_law} does not explicitly guarantee that $F\neq0$ for all $t\geq 0,$ the net force required near any desired equilibrium is not-zero, and thus the net force is not likely to converge to zero in well-posed problems.
In practice, to deal with zero-crossings of the net force, projecting the net force $F$ to $-\epsilon$ and $\epsilon,$ where $\epsilon>0$ is a small number, if $F \in (-\epsilon, \epsilon),$ ensures well-posed computation of \eqref{eq:control_law}.

\end{remark}

\subsubsection{Stability Analysis}
The stability analysis of the closed-loop dynamics with the proposed control law uses the Barbashin-Krasovskii-LaSalle invariance principle, which originally appeared as Theorems 1 and 2 in \cite{lasalle1960some}.
%
In particular, the Barbashin-Krasovskii-LaSalles's invariance principle is used to prove the asymptotic stability of the equilibrium point


In the following, Propositions \ref{prop:equilibriumPoint}, \ref{prop:Omega_def}, \ref{prop:E_def}, and \ref{prop:M_def} define sets used in the stability proof and establish their properties.

\begin{proposition}
    \label{prop:equilibriumPoint}
    Consider the system \eqref{eq:z1_dot}--\eqref{eq:z4_dot},
    The solution
    \begin{align}
        (z_1,z_2,z_3,z_4,u)
            =
                \left( z_{\rmd1}, 0,\matl 0 \\ mg \matr, 0, 0 \right)
        \label{eq:equilibrium_point}
    \end{align}
    is an equilibrium point.
\end{proposition}

\begin{proof}
    Subsituting the $z_1, z_2, z_3, z_4,$ and $u$ given by \eqref{eq:equilibrium_point} in the RHS of \eqref{eq:z1_dot}--\eqref{eq:z4_dot} trivially shows that $\dot{z}_1 = \dot{z}_2 = \dot{z}_3 = \dot{z}_4 = 0.$
\end{proof}

\begin{proposition}
    \label{prop:Omega_def}
    Let $\ell > 0.$ 
    Define
    \begin{align}
        \Omega
            \isdef
                \{
                    z_1,z_2,z_3,z_4,u \in 
                    \mathbb{R}^2,
                    : V
                    \leq \ell
                \}.
    \end{align}
    Then, $\Omega$ is bounded and a positively invariant set.
\end{proposition}

\begin{proof}
    Since $V$ is radially unbounded and $\Omega$ is defined by $V \leq \ell,$ it follows from Theorem $2$ in \cite{lasalle1960some} that $\Omega$ is bounded. 
    Since $\dot{V} \leq 0$ and $V > 0,$ $\Omega$ is a positively invariant set.
\end{proof}

\begin{proposition}
    \label{prop:E_def}
    Define $E \subset \Omega$ such that 
    \begin{align}
        z_1
            &=
            z_{\rmd1}
                    -
                    Q
                    \left(
                        f_2
                        +
                        g_2(z_3)
                    \right), \nn \\
        z_2 &= 0,    \nn \\
        z_4 
            &=
            -
            k_2
            D(\overline{x}_2)
            D({\rm Ch}_{\zeta_1})^4
            \left(
                f_2
                +
                g_2(z_3)
            \right). \nn         
    \end{align}
    %
    Then, $E$ is the set of all points in $\Omega$ where $\dot{V} = 0.$
\end{proposition}

\begin{proof}
    It follows from \eqref{eq:V_dot_final} that $\dot{V} = 0$ if and only if $k_1e_1 = e_2,$ 
    and 
    $e_3 = e_4 = 0.$

    First, $k_1e_1 = e_2,$ in \eqref{eq:e2_def}, necessarily requires that $\tan(\zeta_2) = 0,$ 
    which implies that $z_2 = 0,$ $\zeta_2 = 0,$ $D({\rm Ch}_{\zeta_2}) = I_2,$ where $I_2$ is the $2$ by $2$ identity matrix, and $D({\rm Sh}_{\zeta_2}) = 0,$
    where, for $q \in \mathbb{R}^n,$
    \begin{align}
        \rm Sh_q
            \isdef
                \sinh
                    \left(
                        \matl
                            q_1
                            &
                            q_2
                            &
                            \cdots
                            &
                            q_n
                        \matr^{\rmT}
                    \right)
                    \in \mathbb{R}^n
    \end{align}
    
    Next, \eqref{eq:M_def} with $\zeta_2 = 0$ yields 
    $Q = D({\rm Ch}_{\zeta_1})^2D_{I}(\overline{x}_2)^2,$ 
    and \eqref{eq:e2_dot_def} with $\zeta_2 = 0$ yields
    $\dot{e}_2 = D({\rm Ch_{\zeta_1}})^2\left(f_2 + g_2(z_3)\right)$
    Furthermore, it follows from \eqref{eq:zeta2_ddot} that $\dot{\zeta}_2 = f_2 + g_2(z_3)$

    Therefore, \eqref{eq:e4_def} yields
    \begin{align}
         e_4 
            &= 
                D_{I}({\rm Ch}_{\zeta_1})^2
                D_{I}(\overline{x}_2)
                z_4
        \nn\\
        \quad &
                +
                k_2
                D({\rm Ch}_{\zeta_1})^2
                \left(
                    f_2
                    +
                    g_2(z_3)
                \right).
        \label{eq:e4_stability_analysis}
    \end{align}

    From \eqref{eq:e4_stability_analysis}, $e_4 = 0$ if and only if
    \begin{align}
                z_4
            &=
                -
                k_2
                D(\overline{x}_2)
                D({\rm Ch}_{\zeta_1})^4
                \left(
                    f_2
                    +
                    g_2(z_3)
                \right).
        \label{eq:e4_zero_cond}
    \end{align}
    Next, it follows from \eqref{eq:e3_def} and $e_2 = k_1e_1$ that 
    $e_3 = 0$  if and only if
    \begin{align}
        z_1
            &=
                z_{\rmd1}
                -
                Q
                \left(
                    f_2
                    +
                    g_2(z_3)
                \right),
        \label{eq:e3_zero_cond}
    \end{align}
    which completes the proof. 
\end{proof}

\begin{proposition}
    \label{prop:M_def}
    Let $\SM \subset E$ such that $z_1 = z_{\rmd1},$ $z_3 = \matl0 & mg\matr^{\rm T},$ and $z_4 = u = 0.$ Then, $\SM$ is the largest positively invariant set in $E.$
\end{proposition}

\begin{proof}
    It follows from Proposition \ref{prop:equilibriumPoint} that 
    $
        z_1 
            = 
                z_{\rmd1} 
                    = 
                        D(\overline{x}_1)
                        \atanh(\chi_{\rmd1}),
    $ 
    $z_2 = z_4 = u = 0,$ 
    $z_3 = \matl 0 & mg \matr^{\rm T},$ is an equilibrium point.
    Furthermore, $z_1 = z_{\rmd1},$ 
    $z_2 = z_4 = 0,$
    and
    $z_3 = \matl 0 & mg \matr^{\rm T},$ fulfills equations \eqref{eq:e4_zero_cond} and \eqref{eq:e3_zero_cond}, which implies that the equilibrium point 
    $
        z_1 
            = 
                z_{\rmd1} 
                    = 
                        D(\overline{x}_1)
                        \atanh(\chi_{\rmd1}),
    $ 
    $z_2 = z_4 = u = 0,$ 
    $z_3 = \matl 0 & mg \matr^{\rm T},$
    is a subset of E.
    Thus, $\SM$ is the largest positively invariant set in $E.$
\end{proof}

\begin{theorem}
    \label{thrm:stability}
    Consider the system \eqref{eq:z1_dot}--\eqref{eq:z4_dot}.
    Let $(x_1(0), x_2(0) \in \SSS$ and 
    let $|x_{\rmd1}| <\overline{x}_1).$
    Consider the control law \eqref{eq:control_law},
    where 
    $k_1, k_3, k_4 > 0,$ and $k_2 = k_1\inv.$
    Then,
    \small
    \begin{align}
        \lim_{t\to \infty}z_1(t) 
            &= 
                z_{\rmd1} 
                    = 
                        D(\overline{x}_1)
                        \atanh(\chi_{\rmd1}),
        \label{eq:z1limit}
        \\
        \lim_{t\to \infty}z_2(t) &= 0,
        \label{eq:z2limit}
        \\
        \lim_{t\to \infty}z_3(t) &= \matl 0 & mg \matr^{\rm T},
        \label{eq:z3limit}
        \\
        \lim_{t\to \infty}z_4(t) &= 0,
        \label{eq:z4limit}
        \\
        \lim_{t\to \infty}u(t) &= 0,
        \label{eq:ulimit}
        \\
        \lim_{t\to \infty}x_1(t) &= \matl r_{\rmd1} & r_{\rmd2} \matr^{\rmT},
        \label{eq:x1limit}
    \end{align}
    
\end{theorem}

\begin{proof}
    Consider the set $\Omega$ defined in Proposition \ref{prop:Omega_def} and the set $E$ defined in Proposition \ref{prop:E_def}. 
    It follows from Proposition \ref{prop:M_def} that $\SM$ is the largest positively invariant set in $E.$
    It follows from Theorem 2 in \cite{lasalle1960some} that any solution starting in $\Omega$ converge to $\mathcal{M}$ as $t\to\infty,$ which implies \eqref{eq:z1limit}--\eqref{eq:ulimit}.

    Finally, since  
    $
        \lim_{t\to \infty}z_1(t) 
            = 
                z_{\rmd1} ,
    $ 
    \eqref{eq:z1} implies \eqref{eq:x1limit}.
\end{proof}

\section{Numerical Simulation}
\label{sec:simulations}
In this section, we implement the controller developed in the previous section to follow a desired trajectory while ensuring that the pre-specified position and velocity bounds are not violated. 
In other words, the safety set $\SSS$ is forward invariant.

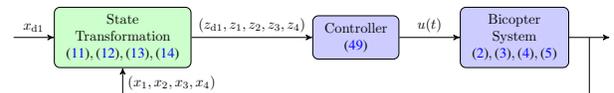
\begin{figure}[!ht]
\centering
    \resizebox{1\columnwidth}{!}{%
    \begin{tikzpicture}[>={stealth'}, line width = 0.25mm]

    \node [input, name=ref]{};
    
    \node [smallblock, fill=green!20, rounded corners, right = 0.5cm of ref , 
           minimum height = 0.6cm, minimum width = 0.7cm] 
           (controller) {$\begin{array}{c} {\rm State} \\ {\rm{Transformation}} \\ \eqref{eq:z1},\eqref{eq:z2},\eqref{eq:z3},\eqref{eq:z4}\end{array}$};
    
    \node [smallblock, rounded corners, right = 3cm of controller, 
           minimum height = 0.6cm , minimum width = 0.7cm] 
           (integrator) {$\begin{array}{c} 
           {\rm Controller} \\ \eqref{eq:control_law}\end{array}$};
    
    \node [smallblock, rounded corners, right = 1.5cm of integrator, 
           minimum height = 0.6cm , minimum width = 0.7cm] 
           (system) {$\begin{array}{c} {\rm Bicopter} \\ {\rm System} \\ \eqref{eq:x1dot},\eqref{eq:x2dot},\eqref{eq:x3dot},\eqref{eq:x4dot}\end{array}$};

    \node [output, right = 1.0cm of system] (output) {};
    
    \node [input, below = 0.75cm of controller] (midpoint2) {};
    \node [input, left = 1.0cm of controller] (reference) {};
    
    \draw [->] (controller) -- node [above] {$(z_{\rmd1},z_1,z_2,z_3,z_4)$} (integrator);
    \draw [->] (integrator) -- node [above] {$u(t)$} (system);
    \draw [->] (system) -- (output);
    \draw [->] (reference.west) -- node [above] {$x_{\rmd1}$} (controller.west);
    
    \draw [-] ([xshift = 0.5cm]system.east) |- (midpoint2);
    \draw [->] (midpoint2) -- node [right] {$(x_1,x_2,x_3,x_4)$} (controller.south);

    \end{tikzpicture}
    }  
    \caption{
        Block diagram illustrating the MCBC controller implementation.
    }
    \label{MCBC_blk_diag}
\end{figure}

The control architecture is shown in Figure \ref{MCBC_blk_diag}

To simulate the bicopter, we assume that the mass of the bicopter $m$ is $1 \ \rm kg,$ its moment of inertia $J$ is $0.2$ $\rm kg  m^2,$ and the length of the bicopter arm $l$ is $0.2$ $\rm m.$
In the controller, 
we set 
$k_1 = 1$, $k_3 = 1$, and $ k_4 = 1. $

The safety set $\SSS$ is specified by
\begin{align}
    \overline{x}_1
        =
            \matl 
                7 \\
                5
            \matr,
    \quad 
    \overline{x}_2
        =
            \matl 
                0.5 \\
                0.5
            \matr,
\end{align}
which implies that the bicopter should stay within a rectangle of width and height $7$ $\rmm$ and $5$ $\rmm$, respectively, and its horizontal and vertical velocity should not exceed $0.5$ $\rm m/s.$

The bicopter is commanded to follow a set of waypoints defined by the vertices of an octagon, which is chosen as the largest octagon within the safe set. 
%
%
The trajectory between the waypoints is constructed using the algorithm described in Appendix A of \cite{spencer2022adaptive} with
a maximum velocity $v_{\rm max} = 1 \ \rm m/s$ and 
a maximum acceleration $a_{\rm max} = 1 \ \rm m/s^2.$ 
The bicopter is commanded to approach the octagon, traverse around the octagon once, and return to the origin.



\begin{figure}[!ht]
    \centering
    \includegraphics[width=1\columnwidth]{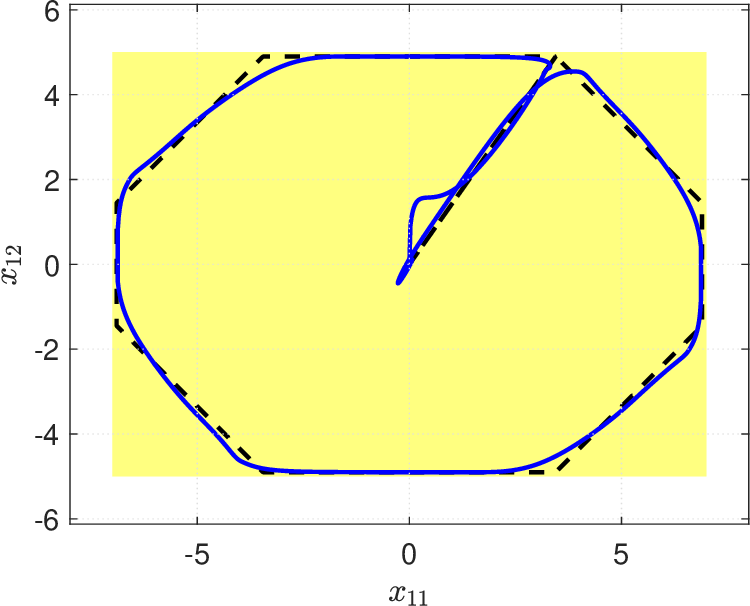}
    \caption{
    \textbf{Constrained trajectory tracking response.}
    Closed-loop position tracking response of the bicopter. 
    The desired position is shown in dashed black, the position response is shown in solid blue, and the safe set is shown in yellow. 
    Note that the position response is always contained within the desired safe set for positions $\SSS.$
    }
    \label{fig:Octogon trajectory}
\end{figure}

\begin{figure}[!ht]
    \centering
    \includegraphics[width=\columnwidth]{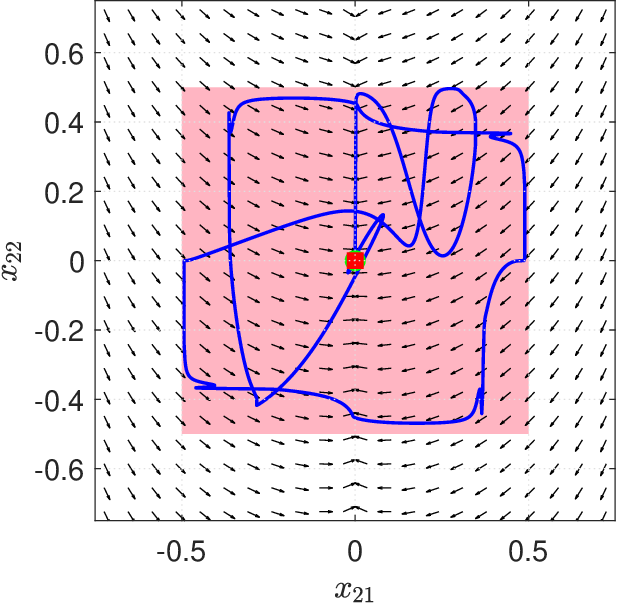}
    \caption{
    \textbf{Constrained trajectory tracking response.}
    Closed-loop velocity response of the bicopter. 
    The velocity response is shown in solid blue, and the safe set is shown in pink. 
    Note that the velocity response is always contained within the desired safe set for velocities $\SSS.$
    }
    \label{fig:Octogon trajectory velocities}
\end{figure}

\begin{figure}[!ht]
    \centering
    \includegraphics[width=\columnwidth]{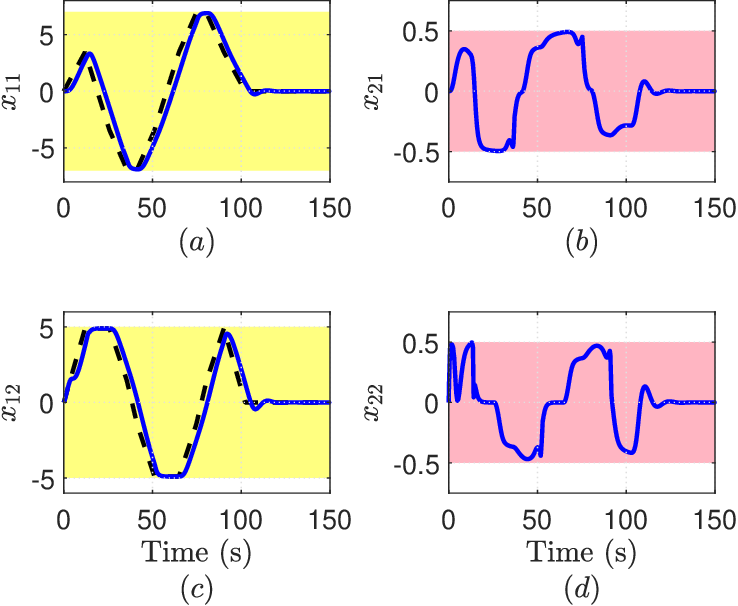}
    \caption{
    \textbf{Constrained trajectory tracking response.}
    Closed-loop position and velocity tracking response with respect to time of the bicopter. 
    Note that the position and velocity responses are always contained within the desired safe set $\SSS.$    
    }
    \label{fig:MCBC_bicopter_pos_vel_states}
\end{figure}

\begin{figure}[!ht]
    \centering
    \includegraphics[width=\columnwidth]{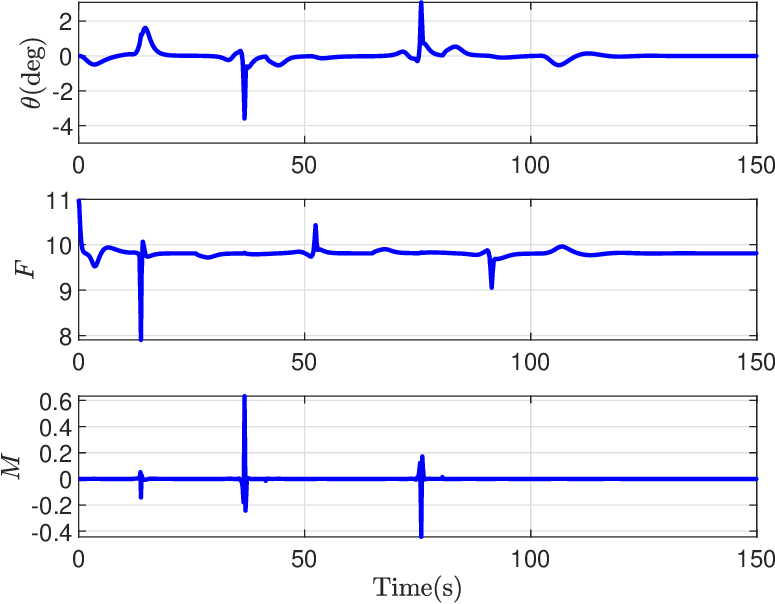}
    \caption{
    \textbf{Octogon Constrained Trajectory:}
    Pitch angle $\theta,$ and inputs $F,$ and $M$ obtained with the control law \eqref{eq:control_law}. Note that the values for the three of them exhibit reasonable values of operation.
    }
    \label{fig:octogon_forces_response}
\end{figure}

Figure \ref{fig:Octogon trajectory} shows the trajectory-tracking response of the bicopter, where the desired trajectory is shown in black dashes, the position response is shown in blue, and the safe set is shown in yellow. 
Figure \ref{fig:Octogon trajectory velocities} shows the velocity response of the bicopter, where the velocity response is shown in solid blue, and the safe set is shown in pink. 
Figure \ref{fig:MCBC_bicopter_pos_vel_states} shows the closed-loop response of the bicopter with respect to time. 
Note that the position and velocity responses are always contained within the desired safe set $\SSS.$   
Finally, Figure \ref{fig:octogon_forces_response} shows the pitch response and the force and the moment applied to the bicopter. 

\section{Conclusions}
\label{sec:conclusions}
This paper presented a novel integrated control law for a bicopter that simultaneously enforced position and velocity constraints while guaranteeing closed-loop stability. 
The approach was based on a state transformation that reformulated the constrained problem as an unconstrained one, employing intermediate Lyapunov candidate functions that grew quadratically near the origin and linearly away from it. 
The proposed controller ensured forward invariance of the safety set, and stability was established using the Barbashin–Krasovskii–LaSalle invariance principle. 
Numerical simulations validated the effectiveness of the controller.
Future work is focused on extending the proposed technique to address asymmetric bounds and parameter adaptation using the results from \cite{portella2025adaptive}.

\printbibliography

\section{Appendix}

\subsection{Math expressions used in the paper}
\label{ap:math_expressions}

It follows from \eqref{eq:zeta_def}, \eqref{eq:z1_dot}, and \eqref{eq:z2_dot} that 
\begin{align}
    \dot{\zeta_1}
        &=
            D_{I}(\overline{x}_1)
             \SF_1(z_1, z_2),
    \label{eq:zeta1_dot}
    \\
    \dot{\zeta_2}
        &=
            D_{I}(\overline{x}_2)
            D({\rm Ch}_{\zeta_2})^2
            \left(
                f_2
                +
                g_2(z_3) 
            \right),
    \label{eq:zeta2_dot}
\end{align}
and thus
\begin{align}
    \ddot{\zeta_1}
        &=
            D_{I}(\overline{x}_1)
            \Big(
                D({\rm Sh}_{2\,\zeta_2})
                D(\dot\zeta_1)
                D(\overline{x}_2)
                \tanh(\zeta_2)
    \nn\\
    \quad &
                +
                D(\rmC_{\zeta_1})^2
                D(\overline{x}_2)
                D_{I}({\rm Ch}_{\zeta_2})^2
                \dot{\zeta_2}
            \Big),
    \label{eq:zeta1_ddot}
    \\
    \ddot{\zeta_2}
        &=
            D_{I}(\overline{x}_2)
            \Big(
                D({\rm Sh}_{2\,\zeta_2})
                D(\dot\zeta_2)
                \left(
                    f_2
                    +
                    g_2(z_3)
                \right)
    \nn\\
    \quad &
                +
                D({\rm Ch}_{\zeta_2})^2
                N(z_3)
                z_4
            \Big).
    \label{eq:zeta2_ddot}
\end{align}

It follows from \eqref{eq:e2_def} that 
\begin{align}
    \dot{e}_2
        &=
            D({\rm Sh}_{2\,\zeta_2})
            D(\dot\zeta_1)
            D(\overline{x}_2)
            \tanh(\zeta_2)
    \nn\\
    \quad &
            +
            D({\rm Ch}_{\zeta_1})^2
            D(\overline{x}_2)
            D_{I}({\rm Ch}_{\zeta_2})^2
            \dot{\zeta}_2,
    \label{eq:e2_dot_def}
\end{align}
and thus
\begin{align}
    \ddot{e}_2
        &=
            2\,
            D(\cosh(2\,\zeta_1))
            D(\dot\zeta_1)^2
            D(\overline{x}_2)
            \tanh(\zeta_2)
    \nn\\
    \quad &
            +
            D({\rm Sh}_{2\,\zeta_2})
            D(\ddot\zeta_1)
            D(\overline{x}_2)
            \tanh(\zeta_2)
    \nn\\
    \quad &
            +
            D({\rm Sh}_{2\,\zeta_2})
            D(\dot\zeta_1)
            D(\overline{x}_2)
            D_{I}({\rm Ch}_{\zeta_2})^2
            \dot{\zeta}_2
    \nn\\
    \quad &
            +
            D({\rm Sh}_{2\,\zeta_2})
            D(\overline{x}_2)
            D_{I}({\rm Ch}_{\zeta_2})^2
            \dot{\zeta}_2
    \nn\\
    \quad &
            -
            2\,
            D({\rm Ch}_{\zeta_1})^2
            D(\overline{x}_2)
            D_{I}({\rm Ch}_{\zeta_2})^2
            D(\tanh(\zeta_2))
            \dot{\zeta}_2
    \nn\\
    \quad &
           +
            D({\rm Ch}_{\zeta_1})^2
            D(\overline{x}_2)
            D_{I}({\rm Ch}_{\zeta_2})^2
            \ddot{\zeta}_2.
    \label{eq:e2_ddot_def}
\end{align}

It follows from \eqref{eq:e3_def} that
\begin{align}
    \dot{e}_3
        &=
            \dot{Q}
            \left(
                f_2
                +
                g_2(z_3)
            \right)
            +
            Q
            N(z_3)
            z_4
            +
            k_2
            \dot{e}_2.
    \label{eq:e3_dot_def}
\end{align}

It follows from \eqref{eq:M_def} that 
\begin{align}
    \dot Q 
        &=
            2
            D({\rm Ch}_{\zeta_2})
            D({\rm Sh}_{\zeta_2})
            D(\dot\zeta_2)
            D_{I}({\rm Ch}_{\zeta_1})^2
            D_{I}(\overline{x}_2)^2
        \nn\\
        \quad &
            -
            2
            D_{I}({\rm Ch}_{\zeta_1})^3
            D({\rm Sh}_{\zeta_1})
            D(\dot\zeta_1)
            D({\rm Ch}_{\zeta_2})^2
            D_{I}(\overline{x}_2)^2,
    \label{eq:M_dot}
\end{align}
and thus
\begin{align}
    \ddot{Q}
        &=
            2
            D({\rm Sh}_{\zeta_2})^2
            D(\dot\zeta_2)^2
            D_{I}({\rm Ch}_{\zeta_1})^2
            D_{I}(\overline{x}_2)^2
    \nn\\
    \quad &
           +
           2
           D({\rm Ch}_{\zeta_2})^2
           D(\dot\zeta_2)^2
           D_{I}({\rm Ch}_{\zeta_1})^2
           D_{I}(\overline{x}_2)^2
    \nn\\
    \quad &
          +
          D({\rm Sh}_{2\,\zeta_2})
          D(\ddot\zeta_2)
          D_{I}({\rm Ch}_{\zeta_1})^2
          D_{I}(\overline{x}_2)^2
    \nn\\
    \quad &
        -
        2
        D({\rm Sh}_{2\,\zeta_2})
        D(\dot\zeta_2)
        D(\tanh(\zeta_1))
        D_{I}(\overline{x}_2{\rm Ch}_{\zeta_1})^2
        D(\dot\zeta_1)
    \nn\\
    \quad &
        +
        6
        D(\tanh(\zeta_1){\rm Sh}_{\zeta_1})
        D_{I}({\rm Ch}_{\zeta_1})^3
        D({\rm Ch}_{\zeta_2}\dot\zeta_1)^2
        D_{I}(\overline{x}_2)^2
    \nn\\
    \quad &
        -
        2
        D_{I}({\rm Ch}_{\zeta_1})^2
        D(\dot\zeta_1)^2
        D({\rm Ch}_{\zeta_2})^2
        D_{I}(\overline{x}_2)^2
    \nn\\
    \quad &
        -
        2
        D_{I}({\rm Ch}_{\zeta_1})^3
        D({\rm Sh}_{\zeta_1})
        D(\ddot\zeta_1)
        D({\rm Ch}_{\zeta_2})^2
        D_{I}(\overline{x}_2)^2
    \nn\\
    \quad &
        -
        2
        D_{I}({\rm Ch}_{\zeta_1})^3
        D({\rm Sh}_{\zeta_1})
        D(\dot\zeta_1)
        D({\rm Sh}_{2\,\zeta_2})
        D(\dot\zeta_2)
        D_{I}(\overline{x}_2)^2,
    \label{eq:M_ddot}
\end{align}

It follows from \eqref{eq:N_def} that
\begin{align}
    N
        &=
            m^{-1}
            \matl
                -\rmC_{z_{31}}z_{32} & -\rmS_{z_{31}}
                \\
                -\rmS_{z_{31}}z_{32} & \rmC_{z_{31}}
            \matr,
    \label{eq:partial_g2_z3}
    \\
    \dot N
        &=
            m^{-1}
            \matl
                \rmS_{z_{31}}z_{41}z_{32} - \rmC_{z_{31}}z_{42} 
                &
                -\rmC_{z_{31}}z_{41}
                \\
                -\rmC_{z_{31}}z_{41}z_{32} - \rmS_{z_{31}}z_{42}
                &
                -\rmS_{z_{31}}z_{41}
            \matr.
    \label{eq:partial_g2_z3_dt}
\end{align}

\end{document}

%% file: PackagesCommands.tex
\usepackage{amsmath} 
\usepackage{amsfonts}
\usepackage{amsthm}

\newtheorem{theorem}{Theorem}

\newtheorem{proposition}{Proposition}[section]
\theoremstyle{definition}

\newtheorem{remark}{Remark}

\usepackage{float} 
\usepackage[skip=0em,font=footnotesize]{caption}

\usepackage[margin = 1in]{geometry}

\usepackage{tikz}
\usetikzlibrary{shapes,arrows,calc,positioning}
\tikzstyle{bigblock} = [draw, fill=blue!20, rectangle, 
    minimum height=6em, minimum width=8em]
\tikzstyle{medblock} = [draw, fill=blue!20, rectangle, 
    minimum height=4em, minimum width=4em]    
\tikzstyle{mux} = [draw, fill=black!20, rectangle, 
    minimum height=5em, minimum width=0.1em]    
\tikzstyle{smallblock} = [draw, fill=blue!20, rectangle, 
    minimum height=2em, minimum width=3em]
    
\tikzstyle{data_block} = [draw, fill=green!20, rectangle, 
    minimum height=2em, minimum width=3em]
\tikzstyle{ops_block} = [draw, fill=blue!20, rectangle, 
    minimum height=2em, minimum width=3em]    
\tikzstyle{est_block} = [draw, fill=red!20, rectangle, 
    minimum height=2em, minimum width=3em]    
    
\tikzstyle{sum} = [draw, fill=blue!20, circle, node distance=1cm,minimum height=0.5cm]
\tikzstyle{signal} = [coordinate]
\tikzstyle{pinstyle} = [pin edge={to-,thin,black}]
\tikzstyle{block} = [draw, fill=blue!20, rectangle, 
    minimum height=3em, minimum width=9em]
\tikzstyle{blockS} = [draw, fill=blue!20, rectangle, 
    minimum height=3em, minimum width=4em]    
\tikzstyle{input} = [coordinate]
\tikzstyle{output} = [coordinate]
\usetikzlibrary{matrix}

\usetikzlibrary{positioning}
\usetikzlibrary{math}

\usepackage{hyperref}
\usepackage{xcolor}
\hypersetup{
    colorlinks,
    linkcolor={blue!100!black},
    citecolor={blue!50!black},
    urlcolor={blue!80!black}
}

\newcommand{\bc}{\begin{center}}
\newcommand{\ec}{\end{center}}
\newcommand{\benum}{\begin{enumerate}}
\newcommand{\eenum}{\end{enumerate}}
\newcommand{\nn}{\nonumber}
\newcommand{\matl}{\left[ \begin{array}}
\newcommand{\matr}{\end{array} \right]}

\renewcommand{\matl}{\begin{bmatrix}}
\renewcommand{\matr}{\end{bmatrix}}

\newcommand{\matls}{\left[ \begin{smallmatrix}}
\newcommand{\matrs}{\end{smallmatrix} \right]}
\newcommand{\isdef}{\stackrel{\triangle}{=}}

\newcommand{\inv}{^{-1}}

\newcommand{\vect}[1]{\overset{\rightharpoonup}{#1}}

\newcommand{\rmA}{{\rm A}}
\newcommand{\rmB}{{\rm B}}
\newcommand{\rmC}{{\rm C}}

\newcommand{\rmI}{{\rm I}}

\newcommand{\rmS}{{\rm S}}
\newcommand{\rmT}{{\rm T}}

\newcommand{\rmc}{{\rm c}}
\newcommand{\rmd}{{\rm d}}

\newcommand{\rmm}{{\rm m}}

\newcommand{\BBR}{{\mathbb R}}

\newcommand{\SF}{{\mathcal F}}

\newcommand{\SM}{{\mathcal M}}

\newcommand{\SSS}{{\mathcal S}}

